\documentclass[aps,prl,twocolumn,10pt,showpacs]{revtex4-1}

\usepackage{graphicx}  
\usepackage{amsmath}

\newcommand{\comment}[1]{}

\usepackage{color}

\definecolor{gray}{RGB}{128,128,128}
\definecolor{green}{RGB}{0,128,0}

\begin{document}

\title{Tunable liquid-liquid critical point in an ionic model of silica}

\author{Erik Lascaris}
\affiliation{Center for Polymer Studies and Department of Physics,
  Boston University, Boston, MA 02215 USA}

\date{8 October 2015}

\begin{abstract}

\noindent
Recently it was shown that the WAC model for liquid silica [L. V. Woodcock, C.
A. Angell, and P. Cheeseman, J. Chem. Phys. {\bf 65}, 1565 (1976)] is
remarkably close to having a liquid-liquid critical point (LLCP).  We
demonstrate that increasing the ion charge separates the global maxima of the
response functions, while reducing the charge smoothly merges them into a LLCP;
a phenomenon that might be experimentally observable with charged colloids.  An
analysis of the Si and O coordination numbers suggests that a sufficiently low
Si/O coordination number ratio is needed to attain a LLCP.

\end{abstract}

\pacs{64.70.Ja, 61.20.Ja}

\maketitle


Tetrahedral liquids tend to display a range of phenomena that are anomalous in
comparison to ``simple'' liquids \cite{ShadrackJabes_JPCM_2012}.  The showcase
example here is liquid water, which displays a large number of anomalies, such
as an increase of the self-diffusion upon compression (diffusion anomaly) and
an increase of the density as it is cooled (density anomaly).  In water, many
of these anomalies are highly pronounced in the supercooled regime, far below
the melting line.  Of particular interest are the seemingly divergent behaviors
of both the isobaric heat capacity $C_P$ \cite{Angell_JPC_1973,
Angell_JPC_1982} and isothermal compressibility $K_T$ \cite{Speedy_JCP_1976}
upon cooling.  Unfortunately these experiments are limited by homogeneous
nucleation, and crystallization rapidly occurs as the temperature approaches
$-40^{\circ}$C \cite{Angell_JPC_1973, Speedy_JCP_1976, Angell_JPC_1982}.

To explain both the anomalies and this divergent behavior, several scenarios
have been proposed \cite{Speedy_JPC_1982mar, Poole_Nat_1992, Sastry_PRE_1996}
among which the liquid-liquid critical point (LLCP) scenario
\cite{Poole_Nat_1992} has received the most attention
\cite{Debenedetti_JPCM_2003, Stanley_BOOK_2013}.  According to this scenario
two metastable liquids exist deep in the supercooled regime: a high-density
liquid phase (HDL) that is highly diffusive, and a low-density liquid phase
(LDL) that is more structured and less diffusive.  These two metastable phases
are separated by a first-order-like liquid-liquid phase transition (LLPT) line
that ends at a critical point.  In the one-phase region beyond any critical
point the response functions remain finite and display a locus of maxima or
minima that near the critical point merges with the locus of correlation length
maxima, known as the Widom line \cite{Xu_PNAS_2005, Stanley_EPJST_2008,
Luo_PRL_2014}.  According to this scenario it is the response function extrema
originating from the LLCP that account for many of the anomalies of water.

The LLCP scenario could also explain the anomalies found in other tetrahedral
liquids.  For example, a LLCP has been found in the Stillinger-Weber model for
liquid silicon \cite{Stillinger_JCP_1974feb, Vasisht_NatPhys_2011}.  Another
candidate is liquid silica, SiO$_2$.  Simulations of the BKS silica model
\cite{vanBeest_PRL_1990} and the WAC silica model \cite{Woodcock_JCP_1976} show
hints of a possible LLCP at low temperatures \cite{Poole_PRL_1997,
SaikaVoivod_PRE_2000, Angell_AIP_2013}, however, more recent studies have
questioned its existence in these models \cite{Lascaris_JCP_2014,
Lascaris_JCP_2015}.  Nonetheless, in the $PT$-plane the isochores of the WAC
model are remarkably close to crossing.  As the crossing of isochores is a
clear indicator of a phase transition \cite{Poole_JPCM_2005,
Lascaris_JCP_2014}, one may therefore conclude that the WAC model is remarkably
close to having a LLCP.

It is important to note that the presence of anomalies does not necessarily
imply the presence of a singularity \cite{Sastry_PRE_1996}, and it is currently
unclear under what exact circumstances a liquid would be able to have a
liquid-liquid transition.  This, together with the fact that LLCPs are
notoriously hard to measure in experiment, has led to an intense debate about
the existence of such a critical point in water \cite{Limmer_JCP_2011,
Sciortino_PCCP_2011, Liu_JCP_2012, Limmer_JCP_2013, Palmer_Nat_2014}, and even
the general existence of liquid-liquid phase transitions in one-component
liquids continues to be questioned \cite{Limmer_JCP_2011}.  It is therefore
important to investigate the conditions under which a LLCP could arise.

Because the WAC silica model is close to criticality it may help us understand
LLCPs in tetrahedral liquids.  In this Letter we modify the WAC model to
include a tunable LLCP.  Silica, as modeled by the WAC model, consists of a 1:2
mixture of Si$^{+4}$ and O$^{-2}$ ions without any explicit bonds.  Apart from
the electrostatic force, the ions also interact with each other via an
exponential term,
\begin{align}
  U_{\text{WAC}}(r_{ij}) = \frac{1}{4 \pi \varepsilon_0} \frac{q_i
    q_j}{r_{ij}} + A_{ij} \exp(-B_{ij} r_{ij})
  \label{EQ:potential_WAC}
\end{align}
Here the subscripts $i,j \in \text{Si,O}$ indicate the species of the two ions
involved, and $q_i$ is the charge of each ion ($q_{\text{Si}}=+4$e,
$q_{\text{O}}=-2$e).  Simulations are performed with $N=1500$ ions and run
for at least $10\,\tau$,
with $\tau$ the approximate equilibration time defined as the average time it
takes for an O-ion to move twice its diameter of 0.28~nm.  Further details of
the implementation, as well as the values of parameters $A_{ij}$ and $B_{ij}$
(which are all positive), can be found in Ref.~\cite{Lascaris_JCP_2014}.

\begin{figure*}[p] 
\centering
\includegraphics[width=0.30\linewidth]{isochores_WAC-108.eps}
\hfill
\includegraphics[width=0.34\linewidth]{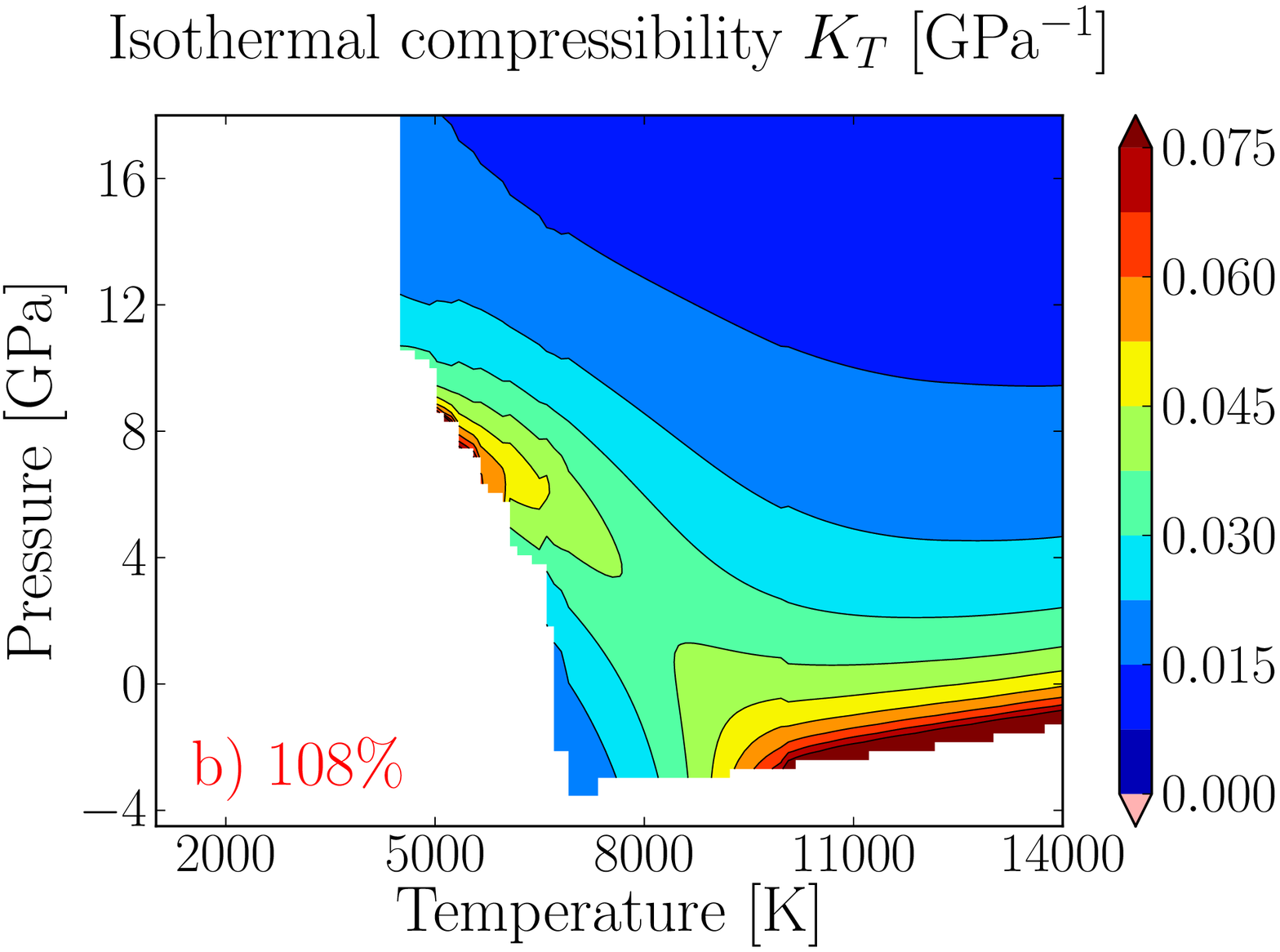}
\hfill
\includegraphics[width=0.34\linewidth]{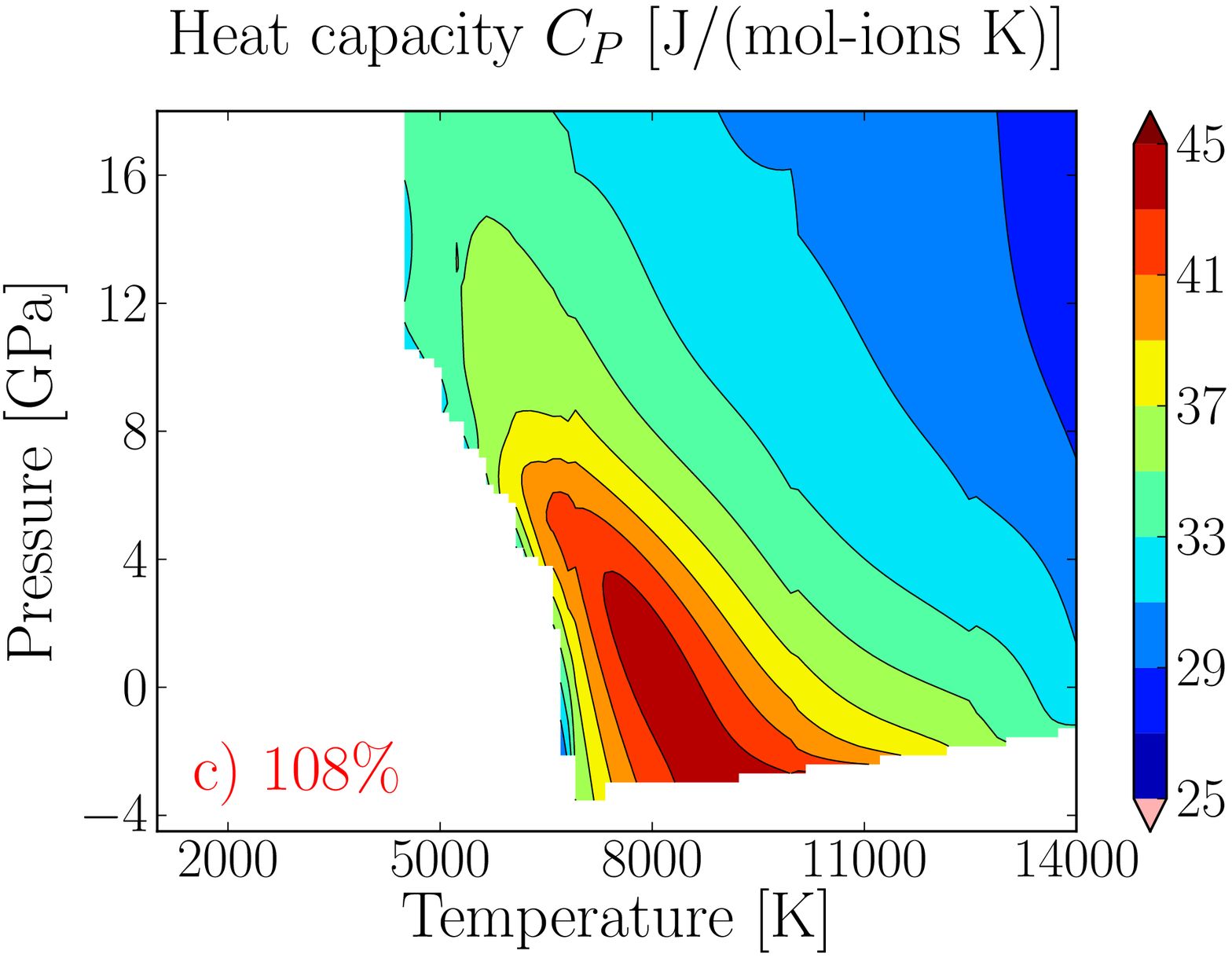}
\quad \\
\quad \\
\includegraphics[width=0.30\linewidth]{isochores_WAC-096.eps}
\hfill
\includegraphics[width=0.34\linewidth]{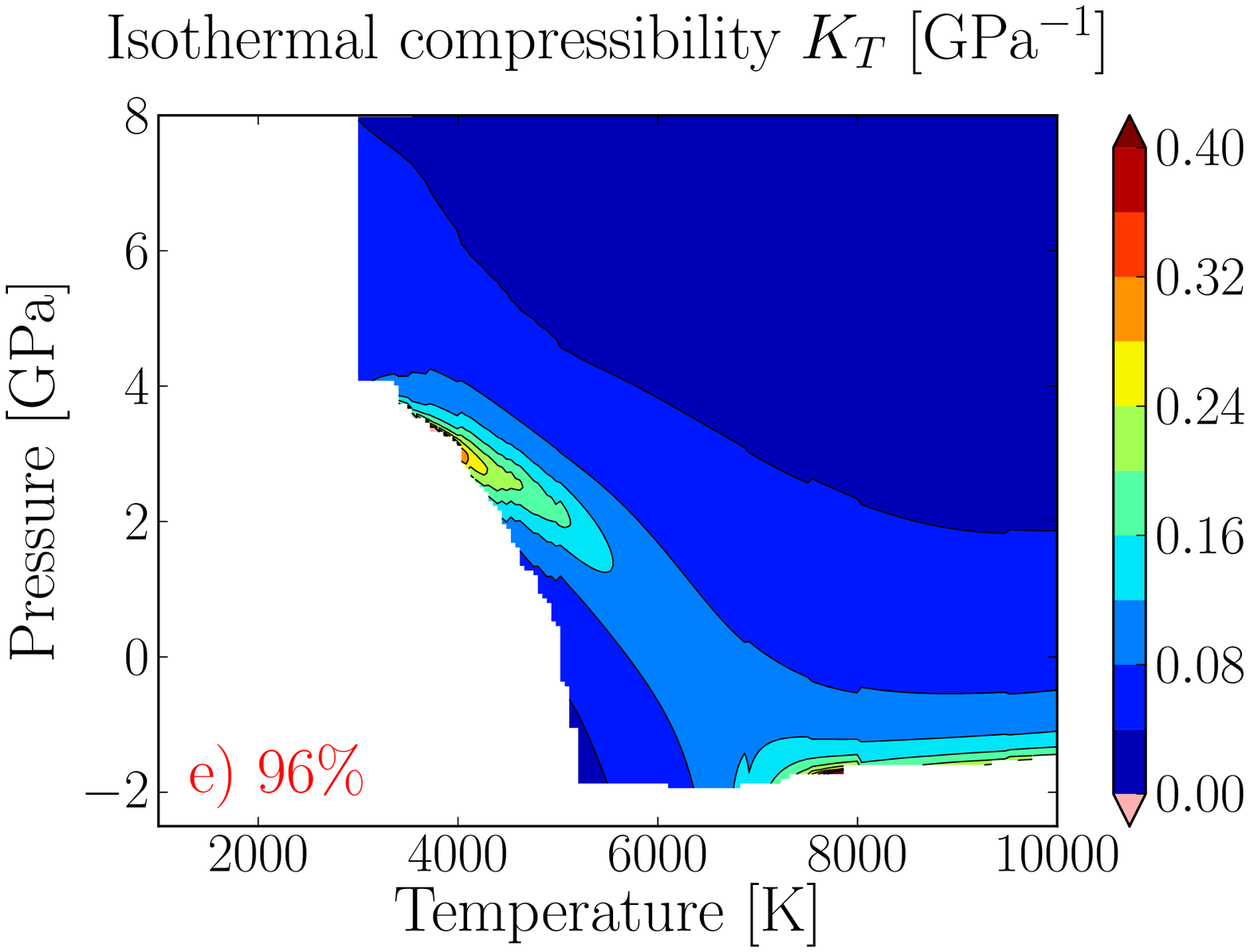}
\hfill
\includegraphics[width=0.34\linewidth]{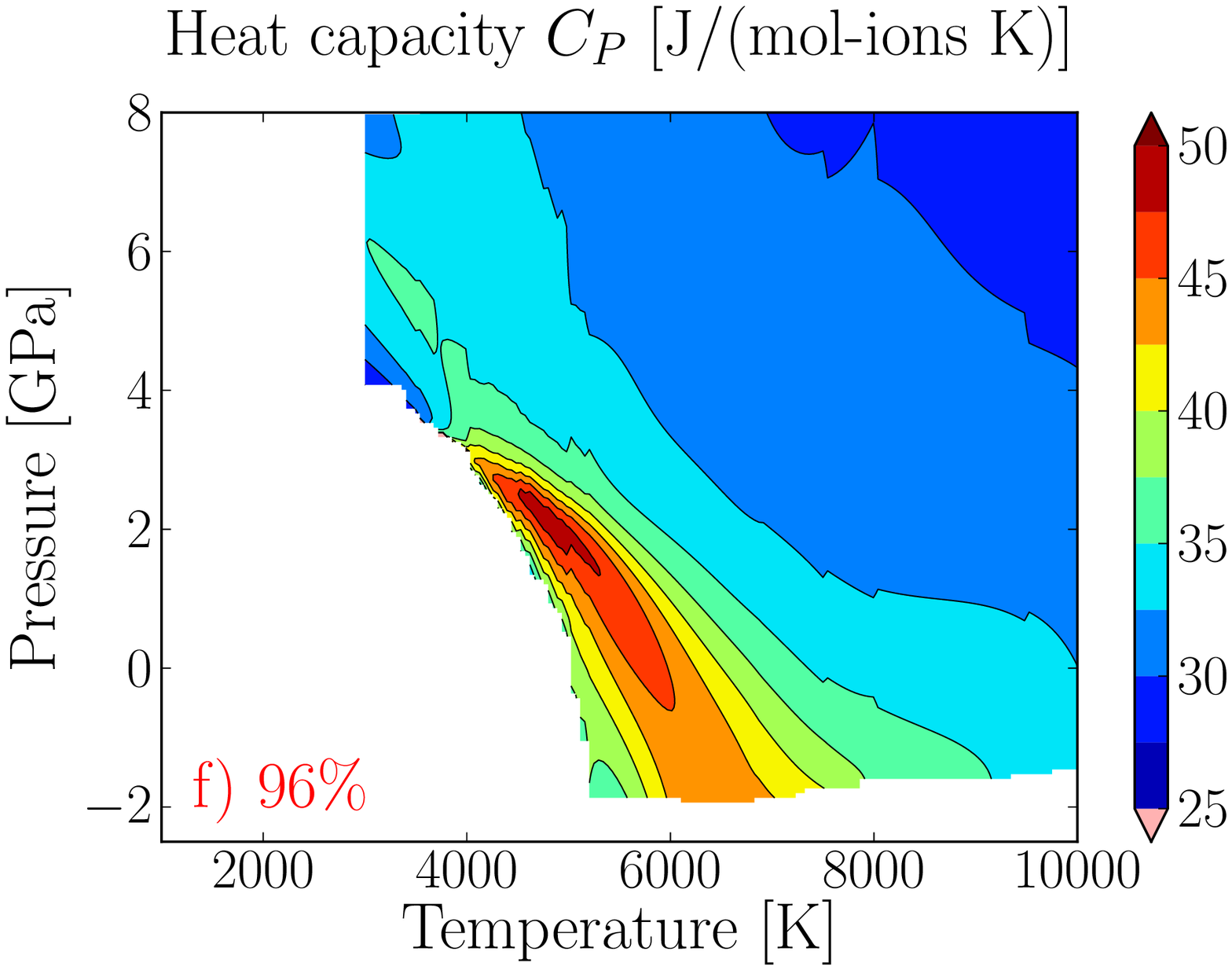}
\quad \\
\quad \\
\includegraphics[width=0.30\linewidth]{isochores_WAC-084.eps}
\hfill
\includegraphics[width=0.34\linewidth]{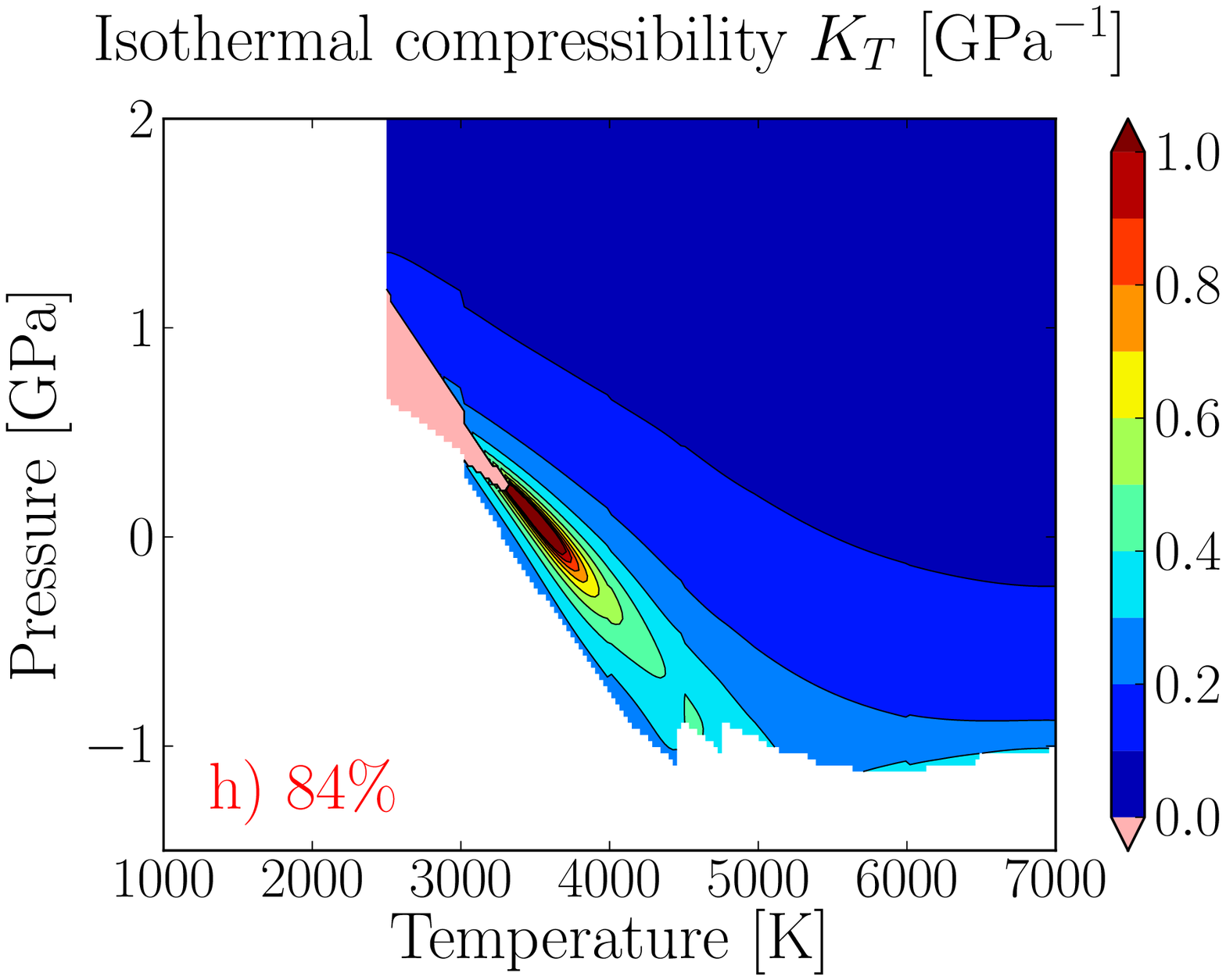}
\hfill
\includegraphics[width=0.34\linewidth]{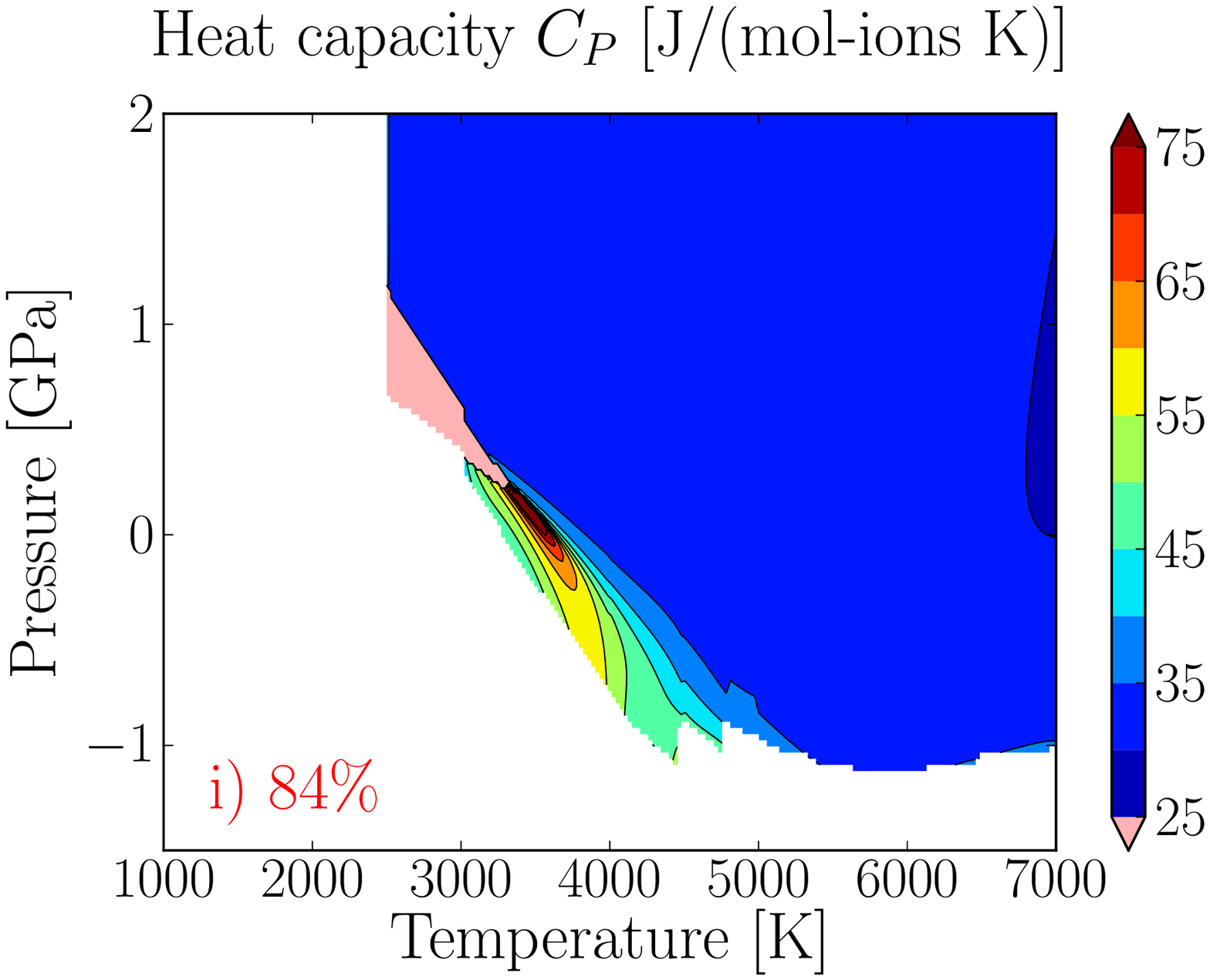}
\quad \\
\quad \\
\includegraphics[width=0.30\linewidth]{isochores_WAC-072.eps}
\hfill
\includegraphics[width=0.34\linewidth]{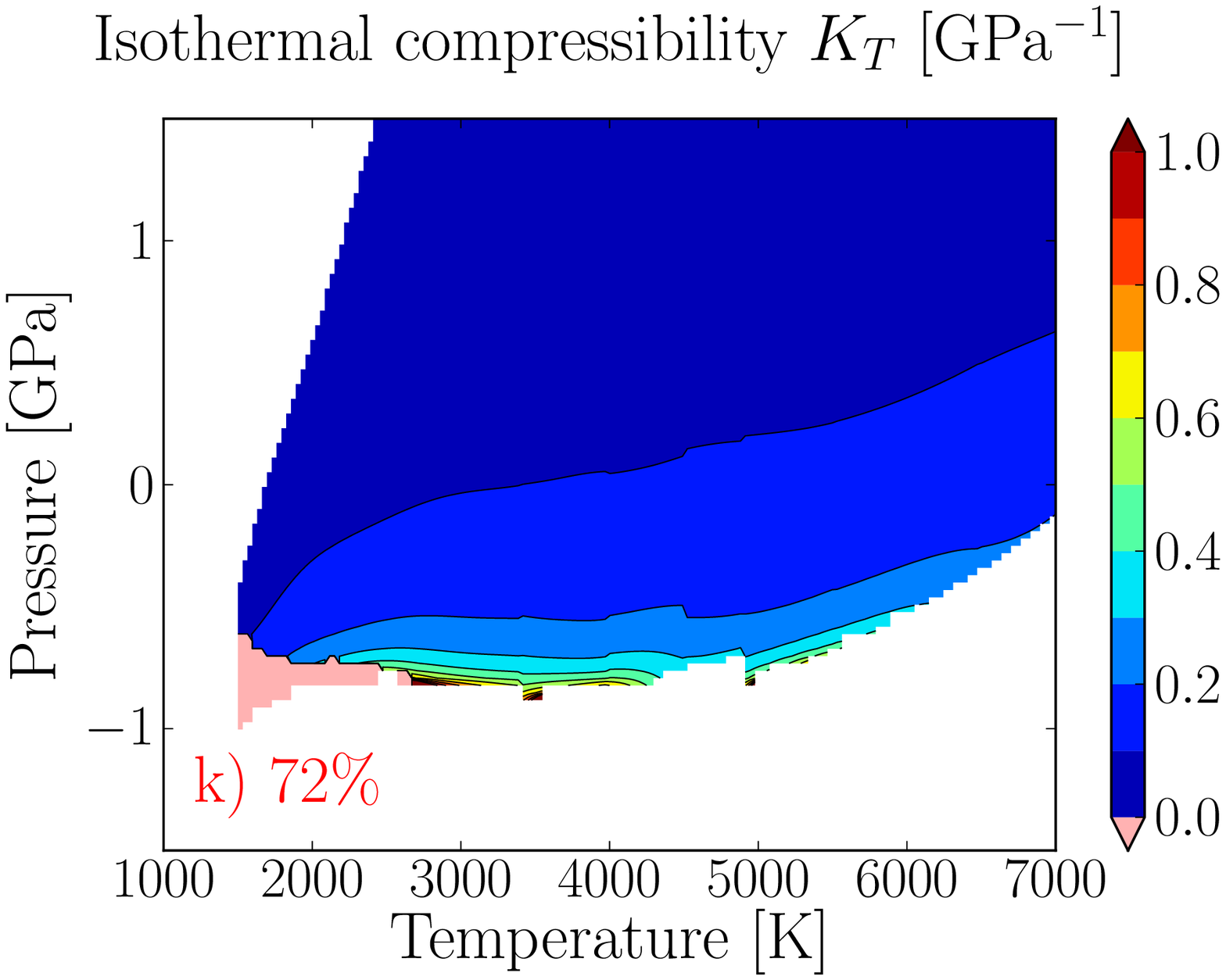}
\hfill
\includegraphics[width=0.34\linewidth]{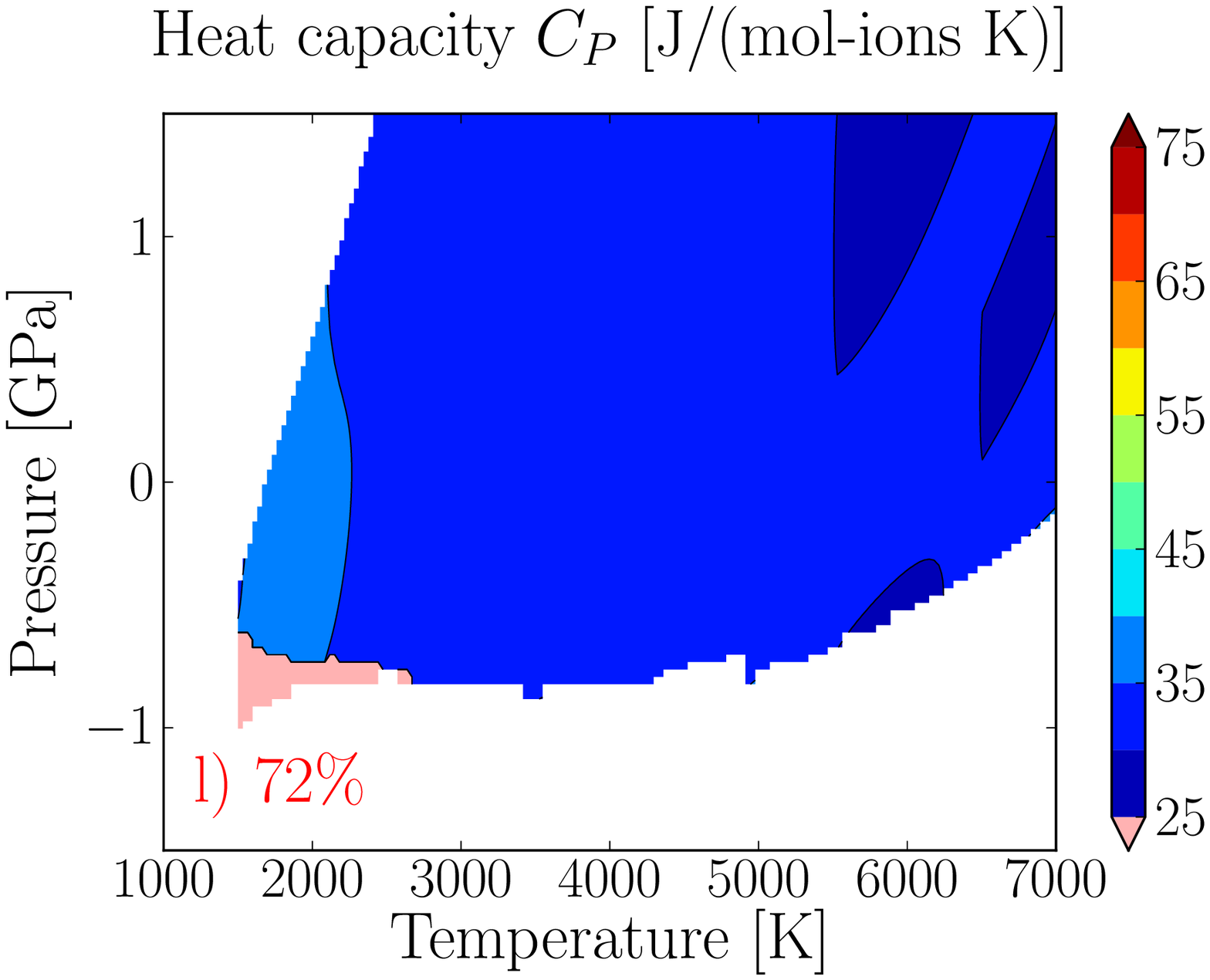}
\caption{
    (color online).  Adjusting the charge of the ions in WAC allows one to
    create or destroy a liquid-liquid critical point (LLCP).  Clear signs of a
    LLCP are (1) the crossing of isochores and (2) when there is a sharp
    increase of the response functions.  In the thermodynamic limit, if a LLCP
    is present then the isochores cross at the same state point as where the
    response functions diverge.  For a finite system, all response functions
    merely show a large maximum near the LLCP.  Left column: isochores, with
    matching colors indicating approximately equivalent isochores (red being
    the isochore that goes through the $K_T$ maximum).  Center and right
    column: isothermal compressibility $K_T$ and isobaric heat capacity $C_P$,
    with the pink area indicating the liquid-liquid coexistence region.  Top
    row: when we increase the ion charge the isochores approach each other but
    do not cross, and both $K_T$ and $C_P$ display a large maximum but at
    different state points.  Second row: upon increasing the charges the
    isochores come closer, while the $K_T$ and $C_P$ maxima start to approach
    each other and grow in magnitude.  Third row: below $f_q \approx 0.94$ the
    isochores cross and the response functions have maxima at the same state
    point: the LLCP.  Bottom row: reducing the charge even further moves the
    LLCP to below the liquid-vapor spinodal.  }
\label{FIG:response_funcs}
\end{figure*}

We modify the WAC model by adjusting the charges by a few percent and keeping
all other parameters unchanged.  In Fig.~\ref{FIG:response_funcs} we consider
the behavior of the isochores and the response functions of the modified WAC
model with 72\% of the original charge ($f_q = 0.72$, i.e.,
$q_{\text{Si}}=+2.88$e and $q_{\text{O}}=-1.44$e), as well as 84\%, 96\%, and
108\%.  All plots are limited at low pressures by the liquid-vapor spinodal,
and at low temperatures by the glass transition line (here taken to be the
point where the oxygen diffusivity drops below $D_{\mathrm{O}} <
10^{-7}$~cm$^2$/s).

In the top row (Figs.~\ref{FIG:response_funcs}a-c) we consider what happens to
WAC when we {\em increase} the charges to 108\% of their original value ($f_q =
1.08$).  The isochores approach each other upon cooling
(Fig.~\ref{FIG:response_funcs}a), which is mirrored by the presence of a large
$K_T$ maximum at the same location in the $PT$-diagram
(Fig.~\ref{FIG:response_funcs}b). Although this would imply the possible
existence of a LLCP at low $T$ around $P = 8$~GPa, consideration of the heat
capacity $C_P$ throws this in doubt because it attains a global maximum at a
significantly different location: near $(T,P) = (8000$~K,~0.7~GPa)
(Fig.~\ref{FIG:response_funcs}c).  If a LLCP were in fact present, all response
functions would diverge upon approaching it.

From Figs.~\ref{FIG:response_funcs}b,c it is not immediately clear that the
$C_P$ maximum and $K_T$ maximum are connected.  However, by reducing the charge
to 96\% ($f_q = 0.96$) we see in Fig.~\ref{FIG:response_funcs}d that the
isochores move closer to one another, that the $K_T$ maximum grows in magnitude
(Fig.~\ref{FIG:response_funcs}e), and that the global $C_P$ maximum moves
towards the $K_T$ maximum (Fig.~\ref{FIG:response_funcs}f).  Ultimately, once
the charge is reduced to below approximately 94\%, a clear LLCP appears.  A
particularly clear example is $f_q = 0.84$, shown in
Figs.~\ref{FIG:response_funcs}g-i, where the isochores cross at the critical
point, and the $C_P$ maximum merges with the $K_T$ maximum at the same state
point.  Reducing the charge further lowers both the critical temperature $T_c$
and the critical pressure $P_c$ until the LLCP disappears below the
liquid-vapor spinodal (Figs.~\ref{FIG:response_funcs}j-l).

The results shown in Fig.~\ref{FIG:response_funcs} raise the obvious question:
why does reducing the charge introduce a LLCP?  Equation~\ref{EQ:potential_WAC}
indicates that reducing the charges makes the Si--O interaction less attractive
and the Si--Si and O--O interactions less repulsive.  Of these competing
effects, the Si--Si is the weakest because its distance is relatively large.
The Si--O interaction is the strongest, and it plays a role analogous to the
hydrogen bond in water.  Consistent with Fig.~\ref{FIG:response_funcs},
reducing the charge reduces the Si--O attraction, which causes an increase of
the volume (decrease of the density) and an increase in diffusivity (i.e., the
glass transition moves to lower $T$).

The competition between the strength of the Si--O bond and the Si--Si bond
becomes clear when we compare the number of neighbors surrounding each Si-ion.
The coordination number $n_{\mathrm{O}}$ is the average number of O-ions
surrounding one Si, and is defined by
\begin{align}
    n_{\mathrm{O}} \equiv 4\pi \rho_{\mathrm{O}} \int_0^{r_{\mathrm{min}}} r^2 g_{\mathrm{SiO}}(r) dr
\end{align}
where $\rho_{\mathrm{O}}$ is the number density of the O-ions,
$g_{\mathrm{SiO}}(r)$ is the Si--O radial distribution function, and
$r_{\mathrm{min}}$ the location of its first minimum.  In the same way we use
$g_{\mathrm{SiSi}}(r)$ an $\rho_{\mathrm{Si}}$ to define the coordination
number $n_{\mathrm{Si}}$ as the average number of Si-ions around one Si-ion.

\begin{figure}[t] 
\centering
\includegraphics[width=1.0\linewidth]{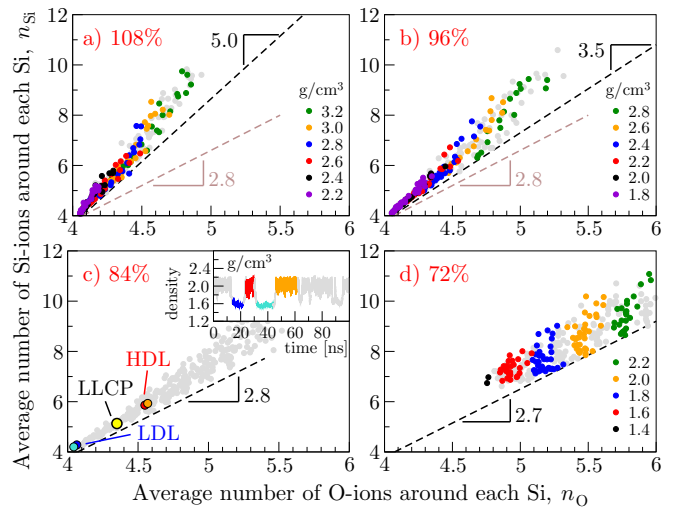}
\caption{
    (color online).  Correlation between coordination numbers $n_{\mathrm{Si}}$
    and $n_{\mathrm{O}}$, for different values of the ion charge.  Colors match
    those of the isochores in Fig.~\ref{FIG:response_funcs}.  At low $T$ the
    liquid prefers the tetrahedral LDL state, surrounded by $n_{\mathrm{O}}
    \approx 4$ O-ions and $n_{\mathrm{Si}} \approx 4$ Si-ions (lower-left
    corner in each panel).  (a,b) Increasing $T$ or $P$ increases both the
    number of O-ions and Si-ions.  However, for each additional O-ion there is
    a minimum number of additional Si-ions, as indicated by the black dashed
    line in each panel.  (c) The slope of this line goes down as the charge is
    reduced, until it goes below 3.5~Si/O and a LLCP appears.  For $f_q=0.84$,
    at (3240~K,~0.30~GPa) the liquid is exactly on the LLPT line and flips
    continuously between LDL (blue, turquoise) and HDL (red, orange), see inset.
    We find that for a LLCP to occur, the HDL phase must have a Si/O
    coordination number ratio below the 3.5~Si/O line.  (d) Reducing the charge
    further makes the LLCP disappear below the liquid-vapor spinodal.  }
\label{FIG:coord-nums}
\end{figure}

Figure~\ref{FIG:coord-nums} shows how the O coordination number correlates with
the Si coordination number.  Silica is a tetrahedral liquid, and therefore at
low $T$ the Si-ions tend to configure with four Si neighbors.  Because of the
Coulomb repulsion, exactly four O-ions are required to act as a ``glue''
between the Si-ions.  Hence, at low $T$ we typically find the liquid near the
$(n_{\mathrm{O}}, n_{\mathrm{Si}})=(4,4)$ state, i.e., the lower-left corner of
each panel in Fig.~\ref{FIG:coord-nums}.  If we increase the temperature or
pressure, a fifth O-ion will move in and produce an imbalance in the charge,
which in turn will attract additional Si-ions.  Increasing $T$ or $P$ thus
increases $n_{\mathrm{O}}$, which leads to an increase in $n_{\mathrm{Si}}$.

The black dashed lines in Fig.~\ref{FIG:coord-nums} represent the minimum
number of Si-ions that surround a cluster of one Si-ion plus $n_{\mathrm{O}}$
O-ions.  The slope of this line depends strongly on the amount of charge that
the ions carry.  The case $f_q=1.08$, shown in Fig.~\ref{FIG:coord-nums}a, has
the most charge per ion and therefore has the largest number of additional
Si-ions per added O-ion: at least 5.0 additional Si-ions for each additional
O-ion (5.0~Si/O).  A reduction in $f_q$ reduces the number of additional
Si-ions per O-ion until below $f_q \approx 0.94$ suddenly a LLCP appears
(yellow circle in Fig.~\ref{FIG:coord-nums}c).  Below $f_q \approx 0.78$ the
liquid-vapor spinodal prevents the formation of a (meta)stable tetrahedral
liquid state (LDL), and the LLCP disappears below the spinodal
(Fig.~\ref{FIG:coord-nums}d).

In Fig.~\ref{FIG:coord-nums}c we focus on the state point (3240~K, 0.3~GPa)
which lies near the LLCP and on the liquid-liquid phase transition line.  This
is clearly demonstrated by the ``phase flipping'' \cite{Kesselring_SR_2012,
Kesselring_JCP_2013} between LDL and HDL that becomes visible when we plot how
the density changes with time (see inset of Fig.~\ref{FIG:coord-nums}c).
Simulations at this state point allow us to compare the properties of the LDL
and HDL phases. As expected, the coordination numbers of LDL lie close to
(4,4), indicating that the liquid is strongly tetrahedral.  The HDL
coordination numbers lie on the opposite side of the LLCP, near
$(n_{\mathrm{O}},n_{\mathrm{Si}}) = (4.6, 6)$, indicating that the average
Si-ion in HDL is surrounded by 0.6 additional O-ions that attract two
additional Si-ions.  Note that the LLCP lies between the LDL and HDL points and
is the average of the two phases.

Fig.~\ref{FIG:coord-nums} suggests that a LLCP is only possible if the HDL has
a Si coordination number below approximately 3.5~Si/O.  Why the coordination
number of HDL matters can be explained using the Gibbs free energy of mixing,
$\Delta G_{\mathrm{mix}} = \Delta H_{\mathrm{mix}} - T \Delta
S_{\mathrm{mix}}$.  We may view the liquid as a mixture of LDL and HDL with
their ratio controlled by a thermodynamic equilibrium, as has been done for
water \cite{Tanaka_EPL_2000, Bertrand_JPCB_2011, Holten_SR_2012,
Tanaka_FD_2013}.  If $\Delta G_{\mathrm{mix}} > 0$ the LDL and HDL will
spontaneously phase separate, and we may witness a liquid-liquid phase
transition.  But if the entropy of mixing $\Delta S_{\mathrm{mix}}$ is large
enough, $\Delta G_{\mathrm{mix}} < 0$ for all temperatures and pressures, and
the liquid will remain homogeneous.  This view together with the results of
Fig.~\ref{FIG:coord-nums} seems to suggest that increasing the charges makes
the Si--O bond stronger, causing a Si ion to draw more Si neighbors into the
first coordination shell.  This then leads to a relative increase in entropy of
the HDL state and an increase of $\Delta S_{\mathrm{mix}}$, with the result
that $\Delta G_{\mathrm{mix}}$ becomes negative for all $T$ and $P$ if the ion
charge is made large enough.  A decrease in ion charge reverses this effect and
allows a liquid-liquid transition to appear.

Values such as the 3.5~Si/O are likely to depend on the parameters of the
model.  We can generalize these ideas, however, by comparing our results to the
idea of ``potential softness'' \cite{Smallenburg_NatPhys_2014}.  A potential
that is too soft, will have too many neighbors per atom in the HDL phase,
leading to an increased entropy of mixing, thus possibly preventing a
transition between LDL and HDL if the Gibbs free energy of mixing becomes
negative for all $T$ and $P$.  Here we adjust the softness by changing the
Si--O strength (via the ion charge), but it is likely that similar findings can
be obtained by careful adjustment of the Van der Waals parameters $A_{ij},
B_{ij}$.

In conclusion, we have shown in this Letter that it is possible for a model to
be tuned such that it smoothly transitions from having a LLCP to not having a
LLCP, and in a manner different from moving the LLCP below a spinodal or glass
transition line.  This means that it is theoretically possible to observe
response function behaviors in a real liquid at high temperatures that seem to
indicate the presence of a LLCP when a LLCP is not actually there, but that
nevertheless the system is ``close'' to having one (as in the case of $f_q \geq
0.96$).  It is therefore important to study a liquid at multiple pressures to
check that there are no separated global response function maxima.  Note that
in the case of water there is experimental evidence that strongly supports the
existence of a LLCP, such as the first-order-like transition between amorphous
ices LDA and HDA \cite{Mishima_Nat_1984, Mishima_Nat_1985, Mishima_JCP_1994}
which would not show hysteresis if water were only ``close'' to having a LLCP.

Finally, this model also presents a possible experimental method of validating
the existence of LLCPs using charged colloids.  Although experiments using
colloids with tetrahedral bonds have already been suggested
\cite{Smallenburg_NatPhys_2014}, our work indicates that a LLCP could also be
obtained by creating a liquid mixture of charged colloids, with the charge
carefully calibrated.

We thank S. V. Buldyrev and F. Sciortino for discussion and comments.  The
Boston University work was supported by DOE Contract DE-AC07-05Id14517, and by
NSF Grants CMMI 1125290 and CHE-1213217.





\end{document}